\documentclass[twocolumn,showpacs]{revtex4}
\usepackage{graphicx}
\usepackage{dcolumn}
\usepackage{amsmath}
\begin{document}
\title{Nonlinear anomalous diffusion equation and
fractal dimension: Exact generalized gaussian  solution}
\author{I. T. Pedron$^1$, R. S. Mendes$^1$,  L. C. Malacarne$^1$, and E. K. Lenzi$^2$}                             %
\affiliation{ $^1$Departamento de F\'\i sica, Universidade
Estadual de Maring\'a, Avenida Colombo 5790, 87020-900
Maring\'a-PR, Brazil\\ $^2$ Centro Brasileiro de Pesquisas F\'\i
sicas, R. Dr. Xavier Sigaud 150,  22290-180 Rio de Janeiro-RJ,
Brazil }

\date{\today}

\begin{abstract}
In this work we incorporate, in a unified way, two anomalous
behaviors, the power law and stretched exponential ones, by
considering the radial dependence of the $N$-dimensional nonlinear
diffusion equation $\partial\rho /\partial{t}={\bf \nabla} \cdot
(K{\bf \nabla} \rho^{\nu})-{\bf \nabla}\cdot(\mu{\bf F}
\rho)-\alpha \rho ,$ where $K=D r^{-\theta}$,  $\nu$, $\theta$,
$\mu$ and $D$ are real parameters and $\alpha$ is a time-dependent
source. This equation unifies  the O'Shaugnessy-Procaccia
anomalous diffusion equation on fractals ($\nu =1$) and  the
spherical anomalous diffusion for porous media ($\theta=0$). An
exact spherical symmetric solution of this nonlinear Fokker-Planck
equation is obtained, leading to a large class of anomalous
behaviors. Stationary solutions for this Fokker-Planck-like
equation are also discussed by introducing an effective potential.
\end{abstract}

 \maketitle

\section{Introduction}

Anomalous diffusion has been the subject of numerous
investigations in the last years. In particular, anomalous
diffusion has played a fundamental role in the analysis  of a wide
class of systems such as diffusion in plasmas \cite{berryman},
turbulent flows \cite{shlesinger}, transport of fluid in porous
media \cite{spohn}, and diffusion on fractals \cite{stephenson}.
In general, anomalous diffusion may be classified employing the
mean square displacement, $\langle (\Delta x)^{2}\rangle $. When
$\langle (\Delta x)^{2}\rangle \propto t^{\eta}$, the value $\eta
>1$ characterizes a superdiffusive process, $\eta <1$ a
subdiffusive one and $\eta =1$ a `normal' diffusion.

We consider here a quite large class of anomalous diffusion,
namely those associated with the $N$-dimensional nonlinear
equation\cite{footnote}
\begin{widetext}
\begin{equation}\label{fok}
\frac{\partial{\rho ({\bf r},t)}}{\partial{t}}= {\bf \nabla} \cdot
\{K{\bf \nabla} [\rho({\bf r},t)]^{\nu}\}-{\bf
\nabla}\cdot[\mu{\bf F}({\bf r},t) \rho ({\bf r},t)]
-\alpha(t)\rho ({\bf r},t),
\end{equation}
\end{widetext}
where $K=D r^{-\theta}$,  $\nu$, $\theta$,  and $D$
are real parameters,  $\mu$ is the drift coefficient, ${\bf
F}({\bf r},t)$ is the external force, and $\alpha (t)$ is a
time-dependent source. From the Gauss theorem one verifies that
this general equation preserves the norm $\int \rho ({\bf r},t)
d^N r$ if $\alpha(t)=0$ and $K{\bf \nabla} [\rho({\bf
r},t)]^{\nu}-\mu{\bf F}({\bf r},t) \rho ({\bf r},t)$ goes
sufficiently rapid to zero when $r=|{\bf x}| \rightarrow \infty$.
The difference between the Fokker-Planck-like equation (\ref{fok})
and the usual one is that the diffusion term depends on a power of
$\rho ({\bf r},t)$ and the diffusion coefficient is spatially
dependent. Notice that this equation for $\theta = 0$ and $\alpha
(t)=0$ has the form:
\begin{equation}\label{porous}
\frac{\partial{\rho ({\bf r},t)}}{\partial{t}}=D {\bf \nabla}
^2[\rho({\bf r},t)]^{\nu}-{\bf \nabla}\cdot[\mu{\bf F}({\bf r},t)
\rho({\bf r},t)],
\end{equation}
and it is usually called porous media equation. The above equation
admits important physical applications such as  percolation of
gases through porous media \cite{muskat}, thin liquid films
spreading under gravity \cite{buckmaster}, some self-organizing
phenomena \cite{carlson}, and surface grow \cite{spohn}, among
others (see Ref. \cite{tsallis1} and references therein).
Furthermore, this equation has been related to nonextensive
Tsallis statistics. This connection was first pointed by Plastino
and Plastino \cite{plastino} and revisited by others authors
\cite{{tsallis1},{ stariolo,compte,compte2,martinez,lisa2,lisa1}}.

Another important kind of anomalous diffusion emerges from Eq.
(\ref{fok}) when $\nu =1$, $\alpha(t)=0$ and ${\bf F}({\bf
r},t)=0$ . In this case, the equation
\begin{equation}\label{richa}
\frac{\partial{\rho ({\bf r},t)}}{\partial{t}}=  {\bf \nabla}
\cdot[K {\bf \nabla} \rho({\bf r},t)]
\end{equation}
was related to turbulent diffusion in the atmosphere
\cite{richardson} (see also Ref. \cite{procaccia2}). Here
$K\propto r^{4/3}$. In general, in a $d$-dimensional space, ${\bf
\nabla} \cdot K{\bf \nabla} $ is proportional to $\frac
{1}{r^{d-1}}\frac{\partial}{\partial
{r}}r^{d-1-\theta}\frac{\partial}{\partial {r}} +
\frac{A}{r^{2+\theta}}$ ($A$ is an operator, which depends on the
angular variables). Thus,
\begin{equation}\label{operador}
\tilde{\Delta}=\frac {1}{r^{d-1}}\frac{\partial}{\partial
{r}}r^{d-1-\theta}\frac{\partial}{\partial {r}}
\end{equation}
is the radial part of the operator ${\bf \nabla}\cdot K{\bf
\nabla}$ to be considered in the study of  $d$-dimensional
spherical symmetrical solutions of Eq. (\ref{richa}). If $d$ is
interpreted as a positive real number,  it plays a role of  a
fractal dimension embedding in a $N$-dimensional space. Thus, Eq.
(\ref{richa}) becomes, in a fractal framework,  the
O'Shaughnessy-Procaccia equation\cite {procaccia},
\begin{equation}\label{frac}
\frac{\partial{\rho (r,t)}}{\partial{t}}=D\tilde{\Delta}
\rho(r,t).
\end{equation}

We are ready now to unify Eqs. (\ref{porous}) and (\ref{frac}). In
fact, if we consider only diffusive terms  and the radial part of
these equations we have \cite{mala}

\begin{equation}\label{unifica}
\frac{\partial{\rho (r,t)}}{\partial{t}}=D\tilde{\Delta}[\rho
(r,t)]^\nu.
\end{equation}

Of course, if $\theta=0$ and  $\nu=1$, the usual diffusion
equation is recovered, {\it i.e.},
\begin{equation}\label{usual}
\frac{\partial{\rho ({\bf r},t)}}{\partial{t}}=D{\bf \nabla} ^2\rho({\bf
r},t).
\end{equation}

In the previous scenario, Eq. (\ref{fok}) can be generalized in
order to be also employed in a fractal context, when spherical
symmetric solutions are considered and ${\bf \nabla} \cdot K {\bf
\nabla}$ is replaced by $\tilde{\Delta}$. This paper is dedicated
to investigate a class of these solutions in the presence of a
source term and with a drift term, as well as the stationary
solutions when a general drift term is present.

The present work is organized as follows. In the next section we
review an  exact solution (generalized gaussian) for Eq.
(\ref{unifica}) and its limit cases. This generalized gaussian
essentially gives the basis for the remaining solutions
investigated here. In Sec. III, a  formal solution for the
nonlinear Fokker-Planck equation Eq. (\ref{unifica}) with a source
term is obtained. Sec. IV deals with the application of our formal
solution for a time power law source. In Sec. V, we present a
generalized gaussian solution for the nonlinear Fokker-Planck with
a linear spherical force term. Furthermore, in Sec. VI, we
consider the stationary solutions for nonlinear Fokker-Planck
equation with a general drift term. A summary of our results is
presented in the last section.

\section{Unified  power-law and stretched exponential solution }

Motivated by the Gaussian solution of the usual diffusion, Eq.
(\ref{usual}), it is natural to employ a generalization of such
solution when we are investigating Eq. (\ref{unifica}). In this
direction, we follow Ref. \cite{mala} employing the ansatz:
\begin{equation}\label{A4}
\rho (r,t)=\left.\left[1-(1-q)\beta(t)
r^\lambda\right]^{1/(1-q)}\right / Z(t)
\end{equation}
if $ 1-(1-q)\beta (t)r^\lambda\geq 0$, and  $\rho (r,t)=0$ if $
1-(1-q)\beta (t)r^\lambda <0$ (cutoff condition). By replacing
this generalized gaussian function in Eq.(\ref{unifica}), we
verify that $\lambda =2+\theta$, $q=2-\nu$,
\begin{equation}\label{A5}
\beta(t)=\beta_0[1+A t]^{-\lambda/[\lambda + d(1-q)]},
\end{equation}
and
\begin{equation}\label{A6}
Z(t)=Z_0[1+ A t]^{d/[\lambda+d(1-q)]},
\end{equation}
where $A=D\lambda (2-q)[\lambda+d(1-q)] \beta_0 Z_0^{q-1}$,
$\beta_0=\beta (0)$ and $Z_0=Z(0)$. It is also important to note
that the product $\beta Z^{\lambda/d}$ is time independent.

From the solution (\ref{A4}) the mean value for a generic function
$f(r)$ can be calculated. In particular, the mean value for
$r^\sigma$ is given by
\begin{equation}\label{A9}
  \langle r^\sigma\rangle=\frac{\int_0^\infty r^\sigma\rho(r,t)r^{d-1}dr}
{\int_0^\infty \rho(r,t)r^{d-1}dr}=C_\sigma \beta
^{-\sigma/\lambda},
\end{equation}
with
\begin{equation}\label{A10}
C_\sigma=\frac{\Gamma(\frac{\sigma
+d}{\lambda})}{\Gamma(\frac{d}{\lambda})}
  \begin{cases}
     \frac{\Gamma(\frac{1}{1-q}+\frac{d}{\lambda}+1)}{(1-q)^{\sigma/\lambda}
     \Gamma(\frac{\sigma+d}{\lambda}
     +\frac{1}{1-q}+1)}&\text{ $q<1$} \\
     \frac{\Gamma(\frac{1}{q-1}-\frac{d+\sigma}{\lambda})}{(q-1)^{\sigma/\lambda}
     \Gamma(\frac{1}{q-1}-\frac{d}{\lambda})}&\text{$\;\, q>1$}.
  \end{cases}
\end{equation}
Observe that the existence of $\langle r^\sigma\rangle$ imposes
restrictions over the parameters: $\lambda+d(1-q)>\sigma(q-1)$ and
$\lambda>0$. In addition, the restriction $q<2$ ($\nu>0$) is
necessary for $\rho(r,t)$ to be real. In the following, we assume
that the parameters obey the above restrictions.

An important particular case of $\langle r^\sigma \rangle$ is the
mean square displacement, $\langle r^2\rangle$, since such mean
value is commonly employed to classify the diffusion process.
Thus, by taking $\sigma=2$ in Eqs. (\ref{A9}) and (\ref{A10}) we
have
\begin{equation}\label{A8}
  \langle r^2\rangle=C_2\beta_0^{-2/\lambda}[1+A
  t]^{2/[\lambda+d(1-q)]}.
  \end{equation}
Consequently, $\langle r^2\rangle\sim t^{2/[2 +\theta+d(\nu-1)]}$
for a sufficient large $t$. Now, it is easy to verify that
$\theta=d(1-\nu)$ describes a `normal' diffusion, even when
$\theta \neq 0$ and $\nu\neq 1$. In fact, there is a competition
between $\theta$ and $\nu$ in such way that the anomalous
diffusive regime induced by $\theta\neq 0$ is compensated by a
convenient one with $\nu\neq 1$. Furthermore, this competition
leads to a subdiffusive process when $\theta>d(1-\nu)$, and to a
superdiffusive one when $\theta<d(1-\nu)$.

Note, also, that an alternative  point of view to verify how much
a solution deviates from the usual one is to analyze the decay of
$\rho(0,t)$. In other words, the way the solution decreases in
this central point can inform about its deviation from the normal
diffusion. In our case, Eq. (\ref{A4}) gives $\rho(0,t)\sim
t^{-d/[2+\theta +d(\nu-1)]}$ in contrast with $\rho(0,t) \sim
t^{-d/2}$, obtained from  the usual diffusion.

We remark that  important limiting cases can be obtained from Eq.
(\ref{A4}).  If $\theta=0$, the solution for the porous media
equation without the drift term is recovered
\cite{tsallis1,plastino},
\begin{equation}\label{media}
  \rho (r,t)=\left.\left[ 1-(1-q)\beta_1(t) r^2\right]
^{\frac{1}{1-q}} \right/Z_1(t),
\end{equation}
with $\beta_1(t)\sim t^{-2/[2+d(1-q)]}$ and $Z_1(t)\sim
t^{d/[2+d(1-q)]}$. On the other hand, for $\nu=1$, the
point-source solution for the O'Shaugnessy-Procaccia
equation\cite{procaccia} emerges,
\begin{equation}\label{fracta}
 \rho (r,t)=\left. e^{ -\beta_2(t)r^{\theta+2}}\right/Z_2(t) ,
\end{equation}
where $\beta_2\sim t^{-1}$ and $Z_2 (t)\sim t^{d/(2+\theta)}$.
Moreover, Eq. (\ref{A4}) recovers the time-translated Gaussian
solution for the Eq. (\ref{usual}),
\begin{equation}\label{usual2}
  \rho (r,t)=\frac{\rho_0}{[4\pi D(t_0+
t)]^{d/2}}~e^{-\frac{r^2}{4D(t_0+t)}}.
\end{equation}
 In fact, Eqs.
(\ref{A5}) and (\ref{A6}) can be respectively written as
\begin{equation}\label{A555}
\beta (t)=\tilde{A}(t_0+t)^ {-\lambda/[\lambda+d(1-q)]}
\end{equation}
and
\begin{equation}\label{A666}
  Z(t)=\tilde{B}(t_0 + t)^{d/[\lambda+d(1-q)]},
\end{equation}
 with $\tilde{A}= \beta_0 A^{-\lambda/[\lambda+d(1-q)]}$,
$\tilde{B}=Z_0 A^{d/[\lambda+d(1-q)]}$, and $t_0=1/A$. Eq.
(\ref{usual2}) is obtained from Eq. (\ref{A4}) with $\beta (t)$
and $Z(t)$ given above in the limit $\nu=1$ and $\lambda=2$, and
by using the normalization $\Omega_d \int \rho (r,t) r^{d-1}
dr=\rho_0$, where $\Omega_d = 2\pi^{d/2}/\Gamma(d/2)$ is the solid
angle in a $d$-dimensional space.

To conclude this review,  we notice that our point-like solution
can be connected with the Tsallis entropy\cite{mala}.

\section{Nonlinear Fokker-Planck equation with time-dependent
source term}

Nonlinear diffusion with absorption arises in many areas of
science such as  engineering, biophysics, solid state physics and
reactor, among others\cite{scott,murray,aris,meyrs}. This section
is dedicated to the analysis of the generalized Fokker-Planck
equation presented in the introduction with a source term. This
kind of equation is expected  to provide a basis for
phenomenological applications, where a large class of nonlinear
anomalous diffusion with  source plays an important role.

Let us consider  the   nonlinear Fokker-Planck equation
(\ref{fok}). The source term in this equation can be removed by an
appropriate change in the solution:
\begin{equation}\label{A1}
\rho ({\bf r},t)= \exp\left( -\int_0^t \alpha(t'){d}t'\right)
\hat{\rho}({\bf r},t).
\end{equation}
In this way, $\hat{\rho}({\bf r},t)$ obeys the equation
\begin{equation}\label{A2}
\frac{\partial {\hat{\rho} ({\bf r},t)}}{\partial{ t}}={\bf
\nabla}\cdot\{\hat{K}{\bf \nabla} [\hat{\rho}({\bf
r},t)]^{\nu}\}-{\bf \nabla} \cdot [\mu{\bf F}({\bf r},t)
\hat{\rho} ({\bf r},t)] ,
\end{equation}
where $\hat{K}(t)=K
\exp\left[{(1-\nu)\int_0^t\alpha(t'){d}t'}\right]$. Thus, Eq.
(\ref{A2}) has the same form of the corresponding one without the
source term, but with a time-dependent diffusion coefficient
$\hat{K}$. Note that this time-dependence is induced by the
nonlinear term $\rho^\nu$, disappearing when $\nu=1$.

As emphasized in the introduction, below Eq. (\ref{usual}), the
replacement of the operator ${\bf \nabla}\cdot (\hat{K}{\bf
\nabla})$ by $\tilde{\Delta}$  unable us also to analyze
situations with fractal dimension in the sense of
O'Shaughnessy-Procaccia. Thus, to investigate a radial  nonlinear
anomalous diffusion equation with a source term and non-integer
dimension $d$, we consider
\begin{equation}\label{source}
\frac{\partial{\rho (r,t)}}{\partial{t}}=D\tilde{\Delta}[\rho
(r,t)]^\nu-\alpha(t)\rho(r,t)
\end{equation}
instead of Eq. (\ref{fok}). In the interests of brevity we omitted
the drift term. The case with an external force is discussed in
Sec. V. To eliminate the source term, we proceed analogously as in
the previous paragraph. Therefore, we obtain
\begin{equation}\label{C1}
 \frac
{\partial{\hat{\rho}(r,t)}}{\partial{t}}=\hat{D}\tilde{\Delta}[\hat{\rho}(r,t)]^{\nu},
\end{equation}
with $ \hat{D}(t)=D
\exp\left[{(1-\nu)\int_0^t\alpha(t'){d}t'}\right]$. Now, we
redefine the time variable introducing an effective time as
\begin{equation}\label{tau}
 \tau(t)=\int_{0}^{t}\hat{D}(t')dt',
\end{equation}
so Eq. (\ref{C1}) can be written as
\begin{equation}\label{A3}
\frac
{\partial{\hat{\rho}(r,\tau)}}{\partial{\tau}}=\tilde{\Delta}[\hat{\rho}(r,\tau)]^{\nu}.
\end{equation}
The solution of this equation was presented and discussed in Sec.
II. Thus, it is given by Eqs. (\ref{A4}), (\ref{A5}) and
(\ref{A6}) with  $D=1$ and $t$ replaced by $\tau$.
\begin{widetext}
Another possible generalized Fokker-Planck is

\begin{equation}
\frac{\partial{[\rho ({\bf r},t)]^\eta}}{\partial{t}}={\bf \nabla}
\cdot\{K{\bf \nabla} [\rho({\bf r},t)]^{\nu}\}  -{\bf \nabla}
 \cdot\{\mu{\bf F}({\bf r},t) [\rho
({\bf r},t)]^\eta\}  - \alpha (t)[\rho ({\bf r},t)]^{\eta}.
\label{fok3}
\end{equation}
\end{widetext}
This equation, with $K$ and $\alpha$ being constants, was
discussed in Ref. \cite{tsallis2}. Note also that Eq. (\ref{fok3})
reduces to Eq. (\ref{fok}) when we replace $[\rho(r,t)]^\eta$ by
$\tilde{\rho}(r,t)$. In the same direction, this kind of
generalization can be incorporated in Eq. (\ref{source}). In this
paper, we will not present details about these possibilities
because their solutions can be directly obtained from Eqs.
(\ref{fok}) or (\ref{source}) by using the identification
$\tilde{\rho}=\rho^\eta$.


\section{Some kinds of sources}

In order to have an enhanced understanding of  the relevance of
the source terms in the anomalous diffusion equation, we apply the
previous general procedure to study some specific cases. In this
context, the simplest situation to be investigated is when
$\alpha(t)=\alpha_0$, where $\alpha_0$ is a constant.  We also
investigate a more generic source term, $\alpha(t)=\alpha_n t^n$.

\subsection{Constant source}

Now, taking into account  Eq.(\ref{source}), we analyze the case
$\alpha(t)=\alpha_0$. Thus, the effective time, Eq. (\ref{tau}),
becomes

\begin{equation}
\tau(t)=D ~ \frac{\exp[(1-\nu)\alpha_0 t]-1}{(1-\nu)\alpha_0}.
\label{B1}
\end{equation}
This effective time exhibits three different behaviors. For
$(1-\nu)\alpha_0<0$, the effective time goes exponentially to
$\tau_{\infty} = D/[(\nu-1)\alpha_0]$, {\it i.e.}, a stabilization
occurs in the diffusive process.  For $(1-\nu)\alpha_0>0$, there
is an exponential increase leading to $\tau(t)\approx D
\exp[(1-\nu)\alpha_0 t]/[(1-\nu)\alpha_0]$, when $t$ is large. It
is interesting to remind that this kind of behavior arises due to
the nonlinearity of the diffusion equation. In this direction, we
have $\tau=Dt$ when $\nu=1$, which corresponds to $q=1$ as long as
$q=2-\nu$.

By using Eqs. (\ref{A5}) and (\ref{A6}) with $t$ replaced by
$\tau$, we obtain
\begin{equation}\label{B2}
  \beta(\tau(t))=\beta_0 \left(1
   + A  ~\frac{\exp[(1-\nu)\alpha_0 t]-1}{(1-\nu)\alpha_0}
  \right)^{-\lambda/[\lambda + d(1-q)]}
\end{equation}
and
\begin{equation}\label{B3}
  Z(\tau(t))=Z_0\left(1+
A ~ \frac{\exp[(1-\nu)\alpha_0 t]-1}{(1-\nu)\alpha_0}
  \right)^{d/[\lambda+d(1-q)]}.
  \end{equation}
Thus, from Eqs.(\ref{A4}) and (\ref{A1}) we verify that
\begin{equation}\label{B4}
\rho(r,t)=e^{-\alpha_0 t} ~ \left. \left[1-(1-q)\beta(\tau(t))
r^\lambda\right]^{1/(1-q)}\right/ Z(\tau(t))
\end{equation}
if $1-(1-q)\beta(\tau(t)) r^\lambda\geq 0$, and $\rho(r,t)=0$ if
$1-(1-q)\beta(\tau(t)) r^\lambda<0$.

In this solution, when $(1-\nu)\alpha_0<0$,  we have, for  large
time,
\begin{equation}\label{max}
\rho(r,t)\approx \left. e^{-\alpha_0 t}\left[1-(1-q)\beta_{\infty}
r^\lambda\right]^{1/(1-q)}\right/Z_{\infty},
\end{equation}
where
\begin{equation}
\beta_{\infty} =\beta_0\left\{1 +
A~/[(\nu-1)\alpha_0]\right\}^{-\lambda/[\lambda+d(1-q)]}
\end{equation}
 and
\begin{equation}
Z_{\infty}= Z_0\left\{1+A ~ /[(\nu-1)\alpha_0]\right\}^{d/[\lambda
+d(1-q)]}.
\end{equation}
This means that $\rho(r,t)$ decays (grows) exponentially in the
time when $\alpha_0>0 ~ (\alpha_0<0)$, and is spatially limited
for $q<1$ as long as $\rho (r,t)=0$ if  $r^\lambda \geq
1/[(1-q)\beta_{\infty}]$.  In contrast, the spatial asymptotic
power law behavior,
\begin{equation} \rho(r,t)\sim
 r^{\lambda/(1-q)},
\end{equation}
arises for sufficiently large $t$  and $(1-\nu)\alpha_0>0$. On the
other hand, when $\nu=1$, we have
 \begin{equation}
\rho(r,t) \sim  t^{-d/\lambda} \exp [-\alpha_0 t -
r^{\lambda}/(D\lambda^2 t)]~
\end{equation}
for large $t$.

Let us focus attention now on the mean square displacement. By
using the effective time in Eq. (\ref{A8}), we obtain
\begin{equation}\label{B2b}
  \langle r^2\rangle=C_2 \beta_0^{-2/\lambda}\left[1+A~
  \frac{e^{(1-\nu)\alpha_0 t}-1}{(1-\nu)\alpha_0}
  \right]^{2/[\lambda+d(1-q)]}.
\end{equation}
The long time behavior for $\langle r^2\rangle$ leads to an
exponential growth when $(1-\nu)\alpha_0>0$. For
$(1-\nu)\alpha_0<0$, this expression converges to an asymptotic
value. For $\nu=1$, we have $\langle r^2\rangle \sim
t^{2/\lambda}$, {\it i.e.},  when $\nu =1$, the mean square
displacement is not affected by the constant source term. These
results show that the presence of a source term plays a relevant
role on the diffusive process when $\nu \neq 1$.

\subsection{Source with a general power term}

Now we  analyze the more general  case corresponding to the source
$\alpha_n(t)=\alpha_n t^n$, where $\alpha_n$ and $n$ are
constants. The related effective time, $\tau_n(t)$, is given by
\begin{equation}\label{B5b}
  \tau_n(t)= \int_{t_0}^t D \exp\left[\frac{(1-\nu)\alpha_n}{n+1}
\left(t^{\prime n+1}-t_0^{n+1}\right)\right] dt' ,
\end{equation}
where  $t_0$ is a cutoff only used to avoid divergence when $n\leq
-1$ and $(1-\nu) \alpha_n/(n+1)>0$. When $n>-1$, we get
\begin{widetext}
\begin{equation}
\tau_n(t)= D\frac{1}{n+1}\left(-\frac
{\alpha_n(1-\nu)}{1+n}\right)^{-1/(1+n)}
\left[\Gamma\left(\frac{1}{1+n}\right) -\Gamma\left(\frac{1}{1+n},
\frac {\alpha_n(\nu-1)t^{1+n}}{1+n}\right) \right]. \label{B5}
\end{equation}
\end{widetext}
Here $\Gamma(c)$ and $\Gamma(c,x)$ are the gamma and  the
incomplete gamma functions, respectively. By using the asymptotic
behavior for $\Gamma(c,x)$, we obtain  $\tau_n \sim t^{-n}
\exp[-\alpha_n(\nu -1) t^{1+n}/(1+n)]$ for $n>-1$ and large $t$.
For $n=-1$, we have $\tau_{-1}(t)\sim t^{\alpha_{-1}(1-\nu)+1}$,
and for $n<-1$, we verify that $\tau_n(t)\sim t$. Taking into
account the asymptotic behavior of the mean square displacement,
$\langle r^2\rangle \sim \tau_n^{2/[\lambda +d(1-q)]}$, we
identify two different behaviors when $n>-1$: if $\alpha_n(\nu-1
)>0$, the mean square displacement reaches a limit value, and  it
grows exponentially if $\alpha_n(\nu-1 )<0$. For $n=-1$, its value
goes with $t^{2 [1-(1-q)\alpha_{-1}]/[\lambda +d(1-q)]}$. Finally,
for $n<-1$, we obtain $\langle r^2\rangle\sim t^{2/[\lambda
+d(1-q)]}$.

\section{External forces}

Now we draw  our attention in order to analyze the generalized
radial Fokker-Planck equation with a drift term. We take Eq.
(\ref{unifica}) with a radial external force, {\it i.e.},
\begin{equation}\label{fe}
\frac{\partial{\rho (r,t)}}{\partial{t}}=D\tilde{\Delta}[\rho
(r,t)]^\nu - \frac{1}{r^{d-1}}\frac{\partial}{\partial r}\left[
r^{d-1} F(r)\rho(r,t)\right],
\end{equation}
where we set $\mu$ equal to unity. In this section we are going to
consider a specific case of a  particle moving in a harmonic
spherical potential $V(r)=  k r^2/2$.  Thus,  $r=0$ is a stable
equilibrium position for positive $k$, and  $r=0$ is an unstable
equilibrium position for negative $k$. Eq. (\ref{fe}), with this
potential, corresponds to the Uhlenbeck-Ornstein
process\cite{Uhlenbeck} in the particular case of $\theta=0$ and
$q=1$ ($\nu=2-q$).

By using, again, the ansatz (\ref{A4}) into Eq. (\ref{fe}), we
verify that $Z(t)$ and $\beta(t)$ obey the equations:

\begin{equation}\label{zt}
  \frac{1}{Z}\frac{dZ}{dt} = Dd\nu\lambda\beta Z^{q-1} -kd
\end{equation}
and
 \begin{equation}\label{bt}
 \frac{1}{\beta}\frac{d\beta}{dt} =-D\nu\lambda^2\beta Z^{q-1}+k\lambda .
\end{equation}
The solution for these two nonlinear coupled equations is
\begin{widetext}
\begin{equation}\label{forca3}
Z(t)=Z_0 \left[ \frac{}{}1+\frac{1}{k}\left( D\nu\lambda \beta_0
Z_0^{q-1}-k \right)
\left(1-e^{-k[\lambda+d(1-q)]t}\right)\right]^{d/[\lambda+d(1-q)]}
\end{equation}
and
\begin{equation}\label{forca4}
\beta(t)=\beta_0\left[1+\frac{1}{k}\left( D\nu\lambda \beta_0
Z_0^{q-1}-k
\right)\left(1-e^{-k[\lambda+d(1-q)]t}\right)\right]^{-\lambda/[\lambda+d(1-q)]},
\end{equation}
\end{widetext}
whose initial conditions are $Z(0)=Z_0$ and $\beta (0)=\beta_0$.

Since $\lambda +d(1-q)>0$, for positive $k$, we have a stationary
solution and $\langle r^2 \rangle$ goes to a constant value. When
$k<0$, there is no stationary state and the $\langle r^2\rangle$
increases exponentially for large $t$. Our solution with
$\theta=0$ corresponds to one obtained by Tsallis and Bukman
\cite{tsallis1} for the one-dimensional case when $\langle x
\rangle=0$. We also remark  that  O'Shaugnessy-Procaccia equation,
Eq. (\ref{frac}), when an external force is incorporated, is a
special case of our results if we take  $\nu=1$.

Before  concluding this section, we note that in the presence of
the source term, $-\alpha (t)\rho$, a similar procedure as
presented in  Sec. III can be employed, leading to Eq. (\ref{fe})
with $D$ replaced by $\hat{D}(t)$. Thus, Eqs. (\ref{zt}) and
(\ref{bt}) must be solved with $\hat{D}(t)$ instead $D$ to obtain
the generalized gaussian solution.

\section{Stationary solutions for the nonlinear Fokker-Planck equation}

We can extend our study about  Fokker-Planck-like equation by
investigating  stationary solutions. The usual Fokker-Planck
equation has a stationary solution that corresponds to the
canonical Boltzmann-Gibbs distribution, $P\propto e^{-\beta V}$.
On the other hand, the corresponding stationary solution of the
nonlinear equation (\ref{porous}) is\cite{lisa1}

\begin{equation}\label{tsallis}
\rho ({\bf r})=\rho_0\left[1-(1-q)\beta V({\bf r})\right
]^{1/(1-q)}
\end{equation}
if $1-(1-q)\beta V({\bf r})\geq 0$, and $\rho({\bf r})=0$ if
$1-(1-q)\beta V({\bf r})<0$. Here $\beta=\mu
\rho_0^{q-1}/[(2-q)D]$, $\rho_0$ is a positive integration
constant, and ${\bf F}({\bf r}) =-{\bf {\bf \nabla}} V({\bf r})$.
Notice that $\beta
> 0$ when $V({\bf r})$ is a confining potential.

When we consider the unified nonlinear equation
\begin{equation}\label{unificada}
\frac {\partial \rho({\bf r},t)}{\partial t}= {\bf {\bf
\nabla}}\cdot \left\{ K({\bf r}){\bf {\bf \nabla}}[\rho({\bf
r},t)] ^\nu \right\}-{\bf {\bf \nabla}}\cdot[\mu({\bf r}) {\bf F}
({\bf r}) \rho({\bf r},t)] ,
\end{equation}
its stationary solution  that generalizes Eq. (\ref{tsallis}) is
\begin{equation}\label{unificada2}
\rho ({\bf
r})=\rho_0\left[1-\left(\frac{1-q}{2-q}\right)\rho_0^{q-1}V_{eff}
({\bf r}) \right]^{1/(1-q)}
\end{equation}
if $1-\left(1-q \right)\rho_0^{q-1}V_{eff}/(2-q)\geq 0$, and $\rho
(r)=0$ if $1-\left(1-q \right)\rho_0^{q-1}V_{eff}/(2-q)<0$, where
we introduced the effective potencial $V_{eff}({\bf r})=\int
\frac{\mu({\bf r})}{K({\bf r})}{\bf {\bf \nabla}} V({\bf r})\cdot
d{\bf r}$. Note that $[\mu({\bf r})/K({\bf r})]{\bf {\bf \nabla}}
V({\bf r})$ must be written as a gradient so it is  possible to
obtain an effective potential  in the $N$-dimensional case. In
particular, for a $3$-dimensional example, this implies that ${\bf
{\bf \nabla}} [\mu({\bf r})/K({\bf r})]\times{\bf {\bf \nabla}}
V({\bf r})=0$. This condition can be easily accomplished if we
take $\mu=\mu(r)$, $K=K(r)$, and $V=V(r)$. Thus,
Eq.(\ref{unificada2}) can be written as
\begin{equation}
\label{unifica3}
 \rho( r)=\rho_0[1-\beta'(1-q)V_{eff} ( r)]^{1/(1-q)},
\end{equation}
with $\beta'=\rho_0^{q-1}/(2-q)$ and $V_{eff}(r)= \int
\frac{\mu(r)}{K(r)}\frac{dV(r)}{d r}~dr$. Following the
development presented in the previous sections, we stress that the
stationary solution (\ref{unifica3}) remains true for an arbitrary
non-integer dimension. For the one-dimensional case, the solution
(\ref{unificada2}) was obtained in Ref. \cite{lisa1}.

\section{Summary}

In this paper we have obtained exact spherically symmetric
solutions for a nonlinear Fokker-Planck equation with
radial-dependent diffusion coefficient, time-dependent source, and
spherical external force. We remark that the time-dependent
solutions are generalizations of the gaussian ones.

Without both external force and source term,   there is a
competition between the fractal aspect and the nonlinear behavior
(dictated by the parameter $\nu$), leading to superdiffusive,
`normal' and subdiffusive processes. On the other hand, in the
presence of a source term, the mean square displacement is only
affected  in the nonlinear case, {\it i.e.}, the source term
modifies  the  superdiffusive, subdiffusive or `normal' characters
of the process when $\nu \neq 1$. In particular, this behavior is
notorious when a power law time-dependent source is considered.

The diffusive process with  a external force $F(r) =- k~r$ leads
to different behaviors to the time dependence of distribution
function. For positive $k$, we have a stationary solution and the
mean square displacement goes to a constant value. When $k<0$
there is no stationary state and the mean square displacement
increases asymptotically in an exponential way.

In the case of the stationary solutions for our Fokker-Planck-like
equation, we obtained a generalization of the canonical
Boltzmann-Gibbs distribution. This generalized solution only
exists if we are able to obtain an effective potential that
includes the true potential, the drift and the diffusion
coefficients.


\end{document}